\documentclass{article}
\usepackage{hyperref}
\usepackage[a4paper, margin=2.5cm]{geometry} % Ajusta los márgenes
\usepackage{enumitem} % Paquete para personalizar la numeración de listas
\usepackage{titlesec} % Paquete para personalizar los formatos de los títulos
\titleformat{\section}[block]{\centering\normalfont\bfseries\Large}{\thesection}{1em}{}
\titleformat{\subsection}[hang]{\normalfont\large\itshape}{\thesubsection}{1em}{}
\renewcommand{\thesection}{\Roman{section}} 
\renewcommand{\thesubsection}{\Alph{subsection}.} % Cambiar numeración de subsección a letras

\usepackage{fontawesome}
\usepackage{placeins}
\usepackage{graphicx}
\usepackage{float}
\usepackage{grffile}
\usepackage{multicol}
\usepackage{lipsum} % Paquete para generar texto de muestra
\usepackage{fancyhdr}
\usepackage{titling}
\title {\fontsize{20}
{28}\selectfont\textbf{\\“MULTIPLE CORRESPONDENCE AND PROPORTIONAL ANALYSIS OF VACCINATION RATE AMONG HEALTHCARE PERSONNEL OF MINSA”}}
\begin{document}
\maketitle

\begin{center}
  \begin{minipage}{0.45\textwidth}
    \centering
    \textit{Valenzuela-Narvaez Luz B.} \\
    \textit{Faculty of Statistic and Computer Engineering} \\
    \textit{Universidad Nacional del Altiplano de Puno,P.O. Box 291} \\
   \textit{ Puno - Peru.} \\
    Email: \textit\texttt{lb.valenzuela@est.unap.edu.pe}
  \end{minipage}
  \hfill
  \begin{minipage}{0.45\textwidth}
    \centering
    \textit{Carlosviza-Amanqui Wladimir A.} \\
    \textit{Faculty of Statistic and Computer Engineering} \\
   \textit{ Universidad Nacional del Altiplano de Puno, P.O. Box 291} \\
   \textit{Puno - Peru. }\\
    Email: \textit\texttt{wcarlosviza@est.unap.edu.pe}
  \end{minipage}

   \begin{minipage}{0.45\textwidth}
    \centering
    \textit{\\Torres-Cruz Fred} \\
    \textit{Faculty of Statistic and Computer Engineering} \\
    \textit{Universidad Nacional del Altiplano de Puno, P.O. Box 291} \\
    \textit{Puno - Peru.} \\
    Email: \textit{\texttt{ftorres@unap.edu.pe}}\\
\end{minipage}
\end{center}

\vspace{10pt}   %espacio
\begin{abstract}
\bfseries
DataProAnalytica is a powerful application for analyzing vaccination data in health care professionals. Through visualizations and multiple correspondence analysis, it uncovers meaningful relationships between variables and categories. The results provide valuable information for improving vaccination strategies. While there are limitations, the potential of DataProAnalytica to improve accuracy and functionality makes it a promising tool for future research and decision making in any other research topic.

.\\
\end{abstract}
\begin{multicols}{2} % Inicio de las dos columnas

\section{Introduction}
In December 2019, a neurometabolic outbreak was detected in Wuhan, China, linked to a wholesale seafood market. The pathogen responsible turned out to be a new coronavirus, initially designated 2019-nCoV by the WHO but later renamed
 
2019-nCoV. The name was changed to SARS-CoV-2, and the disease was renamed COVID-  19\cite{bonilla-sepulveda2020}.
The development of effective vaccines became a crucial challenge to curb the spread of the disease and prevent hospitalizations. This has prompted an accelerated effort in the manufacture and distribution of safe and effective vaccines against COVID-19\cite{verdu2022adherencia}.\\
Therefore, major public and private laboratories have entered a race to find an effective vaccine against it. A vaccine that is capable of generating immunity is the only tool that can slow the spread of the virus. Because the disease is spreading rapidly worldwide, causing a large number of deaths in countries with this disease\cite{gonzalez2020vacuna}.\\
In November 2021, acordtingo data from John Hopkins University \& Medical, m o r e t h a n 258 million confirmed cases and m o r e more than 5 million deaths from COVID-19 were reported worldwide. The Ministry of Health has registered more than 2.2 million confirmed cases, m o r e t h a n 201,000 deaths and more than 91,000 recovered \cite{arpasi2022personal}\cite{cordova2020covid} .\\
The Peruvian government has implemented health policies similar to those in China, such as quarantines and social distancing, to cope with the pandemic\cite{abuabara2020ataque}. Strengthening the health system was also a priority. However, vaccination coverage among health professionals in the country is low, despite their exposure and risk\cite{santacruz2016percepcion}. It is essential to ensure that these professionals receive the necessary biosafety equipment to protect themselves while caring for infected patientsts\cite{cano2022factores}. In Peru, the Occupational Safety and Health Services have the responsibility to protect and promote the safety and well-being of workers. Although there is a shortage of professionals in this field, OSHSS has adapted to the current challenges. According to the ICOH, the coverage of these services worldwide was only 5 percent of the workforce in 2005\cite{guanche2020covid}.\\

It is essential that health care workers have sound knowledge in the prevention of COVID-
19 transmission and access to adequate personal protective equipment, despite limited financial resources\cite{cruz2020protegiendo}. The major challenge lies in the effective training of these professionals to ensure their protection and the safety of patients. Despite the fact that millions of people stay at home to minimize transmission of the virus, health care workers are exposed to a high risk of contracting COVID-19 when caring for patients in hospitals and health centers. Experiences in China and Italy have shown that 20\% of healthcare workers became infected, resulting in the unfortunate loss of life\cite{guanche2020covid}.\\
Evaluation of the impact of COVID-19 vaccination on health care personnel is vital because of their high risk of exposure to the virus. Vaccination can reduce the spread of the virus to the general population. It is important to analyze the acceptance of the vaccine by health care personnel and to understand the factors that influence their decision to vaccinate. This to identify barriers and promote strategies for commodization. This study aimed to identify barriers and promote strategies for coimmunization.

In addition, the effect of vaccination on the incidence of COVID-19 cases among health care providers
sonnel and its impact on the spread of the virus in health care settings can be assessed. A review of the scientific literature can also provide information on adherence to vaccination and factors influencing acceptance or hesitancy of COVID-19 vaccination among health care personnel\cite{verdu2022adherencia}.\\
For this purpose, a multiple correspondence analysis was performed to examine the relationship between variables and categories related to health personnel during the COVID-19 pandemic. Studies in English and Spanish with specific keywords were used. Data were obtained from the open database of the Peruvian Ministry of Health \cite{web_page}. The objective is to identify significant patterns and relationships to better understand the situation of health professionals during the pandemic. This analysis will provide accurate information for decision making and implementation of appropriate protective measures.
In this article, the background of the question based on the acceptance of vaccines to counteract the infections of this disease by health professionals was demonstrated in the background. In such a way, it was to regain institutional trust so that the necessary levels of collective immunity against COVID-19 could be achieved through mass and voluntary vaccination of citizens, thus having a better chance of fighting this disease with our health professionals\cite{gonzalez2020vacuna}.

%METODOLOGIA
\section{Methodology}

\subsection{Purpose and type of study}
The objective of this scientific article is to perform a systematic evaluation of the vaccination rate in health professionals of the Ministry of Health (MINSA) from the beginning of the pandemic to the present. The research will focus on analyzing the acceptance of the COVID-19 vaccine among health personnel and understanding the factors that influence their decision to be vaccinated. To this end, a multiple correspondence analysis will be carried out, exploring the relationship between vaccination and the various associated factors. The search for information will be considered in all departments of the country, including their respective DIRESA. The key words used in the search will be COVID-19, analysis multiple ,  health personnel and relationship.
\subsection{ Multiple Correspondence Analysis}
Multiple correspondence analysis (MCA) is a technique used by Bourdieu and his team to analyze the relationship between theory and empirical evidence. In this article, the MCA was applied to examine the configuration of the private university space in Argentina (1955-1983), identifying the relationships of similarity and difference between institutions. The MCA made it possible to graphically represent these relationships, revealing the structure and dynamics in the visualization of the data with respect to health personnel and the distinctive characteristics of each institution to which they belong\cite{a}.
\subsection{Techniques and instruments}
The data visualization technique was used with the ACM method to systematize the evaluations carried out on vaccines to health professionals; a data sheet was elaborated to record the information. The indicators used were health professionals and the type of manufacturer, taking into account the comparison between the positive relationships between categories and variables, represented by means of graphs, providing the visualization of the relation-ships that can be analyzed.
\subsection{Data collection and processing}

The COVID-19 vaccine database was obtained from the National Open Data Platform, specifically from the dataset provided by the Ministry of Health (MINSA) regarding COVID-19 vaccination. The focus of the analysis was on health professionals. Initially, we had an original database that included 1,048,576 records with several variables, such as cutoff date, UUID, risk group, age, sex, date of vaccination, dose, manufacturer, Diresa, departure, province, district, and type of age. To focus on health professionals, a filter was performed on the original database. In particular, the records that had the variable “risk group” associated with health professionals were selected. In this way, a subset of data was obtained that corresponded to vaccinated health professionals. Subsequently, the relevant variables for the analysis, such as age, sex and dose, were selected. These variables would provide key information on health professionals in relation to COVID-19 vaccination. The collected and filtered data are available at the following link: Vaccination against COVID-19 - Ministry of Health
- MINSA\cite{web_page}.\\\\

\begin{figure}[H]
  \centering
  \includegraphics[width=0.5\textwidth]{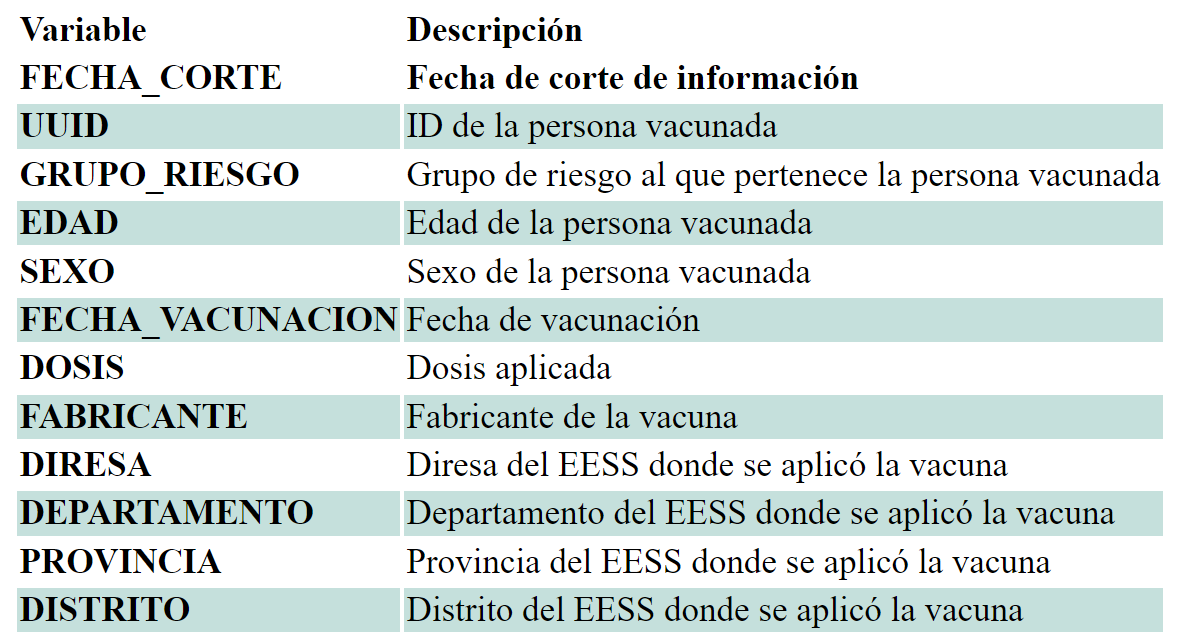}
  \caption{Data dictionary}
  \label{fig:o}
\end{figure}

\subsection{Exploratory analysis}
The sample used in the analysis was 3,915 participants, which provided a solid database to obtain representative and reliable results. Several key variables were considered in the analysis, including sex, age and dose administered. The sex variable was divided into two categories: male and female. The age variable was recorded as a continuous variable and was used to identify possible patterns and differences in specific age groups. In addition, information was collected on the dose administered to each health professional. The main variable of interest in this analysis was membership in a specific group of health professionals. This variable was divided into three categories: health science students, health science interns, and health personnel. These categories allowed differentiation between different levels of experience and roles within the health care setting.
Using the method of correspondence analysis, we sought to identify patterns, relationships and associations between these variables and categories. The results were presented in a reduced dimension space, which facilitated a clear and concise visualization of the relationships between the categories and the variables studied.
\subsection{Logical implementation}
Code in the R language will be developed. \\\\
\end{multicols} % Fin de las dos columnas
\begin{figure}[H]
  \centering
  \includegraphics[width=0.8\textwidth]{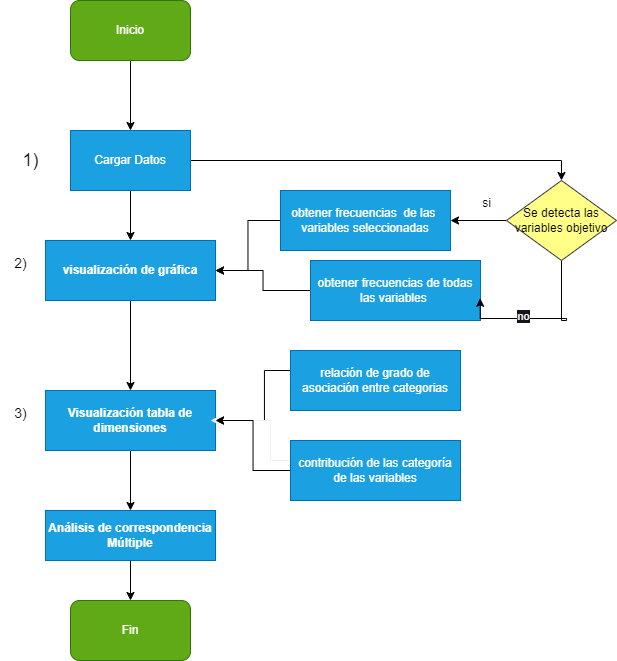}
  \caption{Flowchart}
  \label{fig:o}
\end{figure}

\begin{multicols}{2} % Inicio de las dos columnas
1.-The Vacuum Indicators data is loaded in xlsx format, the code is in R language and is executed in the R script part.\\\\
2.-The code begins to read and count the variables of interest. It shows descriptions of the first two dimensions obtained in the multiple correspondence analysis.\\\\
3.-Shows the relationship between the categorical variables of the categorical variables and includes a table of vaccination rate by risk group. Finally, it shows the cos2 of the individuals for the 2 dimensions, grouping of individuals and visualization of ellipses.
\subsection{Implementation}
The following libraries: \\\\
Shiny library: This is the main Shiny library that allows interactive web applications to be created in R.\\\\
The dplyr library is a library used for data manipulation, such as filtering, grouping and re summarizing data.\\\\
The caret library provides tools for the training and evaluation of automatic learning models.\\\\
The tidyr library: Used for data manipulation and cleaning, especially for transforming wide-format data to long-format data.\\\\
The factoextra library provides functions for vi- sualizing and analyzing multiple correspondence analysis (MCA).\\\\
Separate variables for the columns risk group, age, sex and manufacturer are created from the selected data. Use the cut function to create intervals for the age variable.\\\\
Creates a data frame for each variable with the cate- gories and their corresponding frequencies. Plot the bars for each variable: The visualization functions of ggplot2 are used to create bar graphs representing the frequencies of each category of the variables risk group, age, sex and manufacturer.\\\\
The MCA function of the FactoMineR package is used, resuming the multiple correspondence analysis using the data variables.\\\\

\end{multicols} % Fin de las dos columnas

\section{Results}

\begin{figure}[H]
  \centering
  \includegraphics[width=0.8\textwidth]{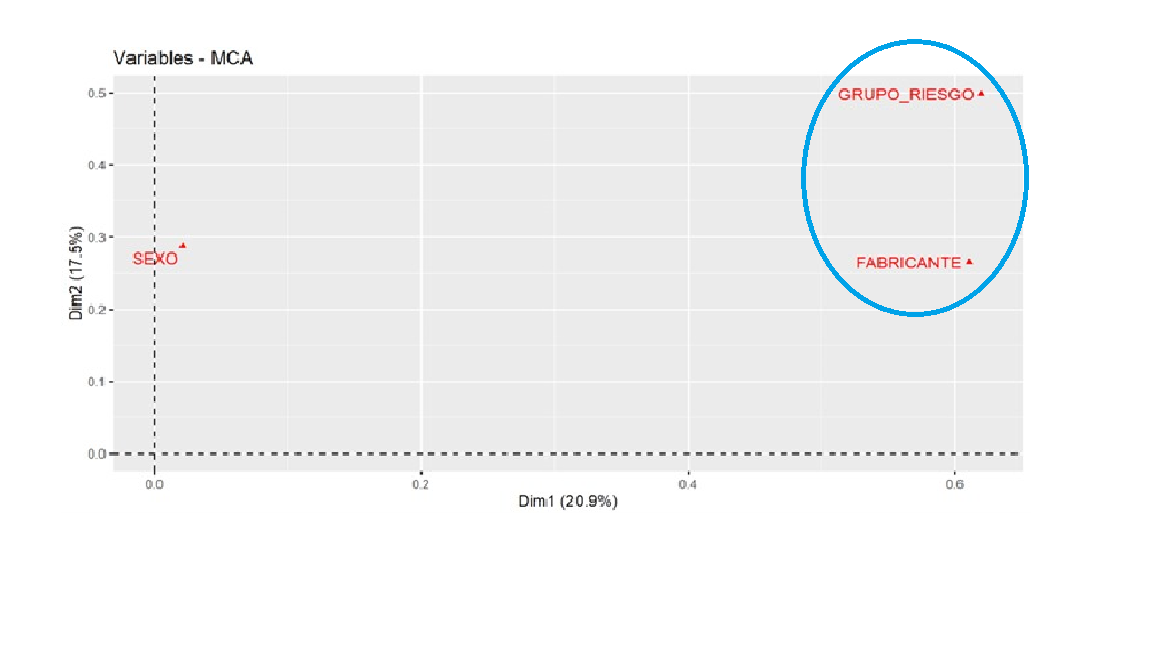}
  \caption{Point cloud between variables}
  \label{fig:Imagen1}
\end{figure}

The results of the work were obtained from 3,915 data points, with the variables described in the methodological section, where it is notably shown that at the moment of identifying the variables, these are correlated with each dimension. The squared correlations between variables and dimensions are used as coordinates. In this case, the Risk Group variable is the one that presents the highest correlation with dimension 2, by a small difference with sex, which is slightly lower. Likewise, the variable m o s t correlated with dimension 1 is manufacturer.
\autoref{fig:Imagen1}.

\begin{figure}[H]
  \centering
  \includegraphics[width=0.8\textwidth]{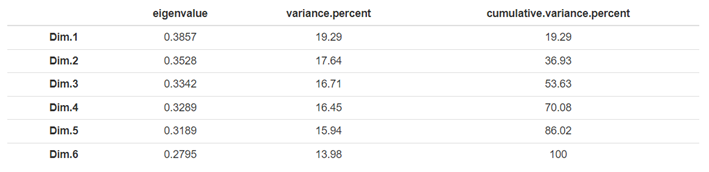}
  \caption{Table of percentage variances}
  \label{fig:Im}
\end{figure}

The following table shows how the percentage of variance accumulates according to the explanatory profile as 6 dimensions are considered in the multiple correspondence analysis.\autoref{fig:Im}\\

\begin{figure}[H]
  \centering
  \includegraphics[width=0.8\textwidth]{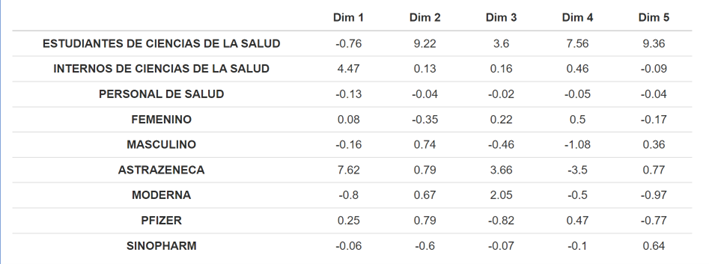}
  \caption{Dimension Table}
  \label{fig:Imagen2}
\end{figure}

\begin{figure}[H]
  \centering
  \includegraphics[width=0.8\textwidth]{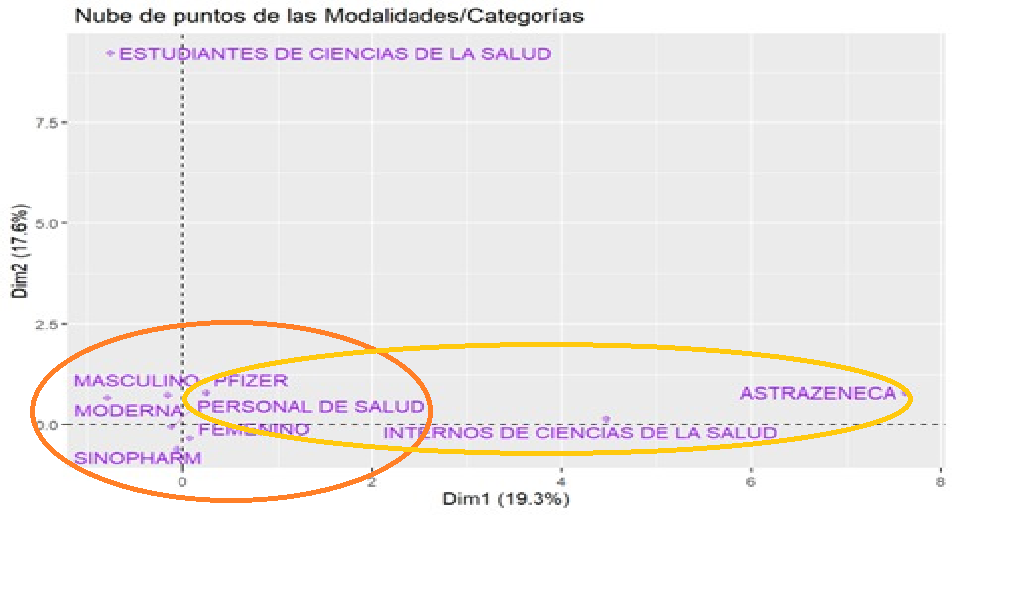}
  \caption{Point cloud of the Modalities/Categories}
  \label{fig:Imagen3}
\end{figure}

The following table shows the point coordinates of each category in each dimension . \autoref{fig:Imagen2}.\\\\
The following figure shows the relationship and association between the categories of the variables and can be interpreted as follows: Categories of variables with a similar profile are grouped together. The categories of negatively correlated variables are positioned on opposite sides of the origin of the graph (opposite quadrants). The graph shows that it is easier to observe the categories that are better represented by dimension 1 than the health sciences interns, health personnel and AstraZeneca vaccine, which present a better profile. 
\autoref{fig:Imagen3}
\\\

\begin{figure}[H]
  \centering
  \includegraphics[width=0.8\textwidth]{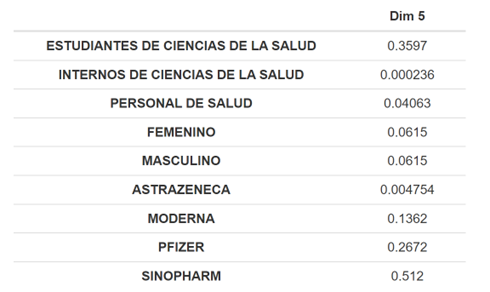}
  \caption{Table of dimensions}
  \label{fig:Imagen5}
\end{figure}

In the following figure, it can be seen that none of the categories is well represented by 2 dimensions, and the modern vaccine category has a cos2 of 0.6138, which is not close enough to 1. All the variable categories would require m or e t h a n one dimension to be better represented. \autoref{fig:Imagen5}\\
\begin{figure}[H]
  \centering
  \includegraphics[width=0.8\textwidth]{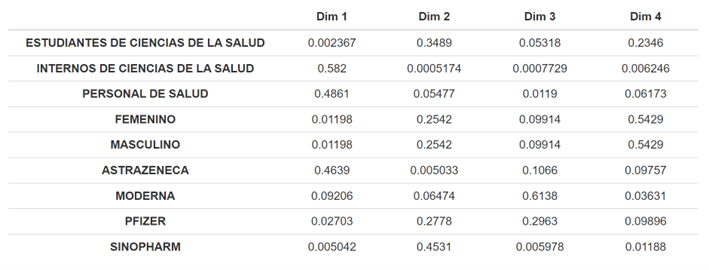}
  \label{fig:Imagen4}
\end{figure}

\begin{figure}[H]
  \centering
  \includegraphics[width=0.8\textwidth]{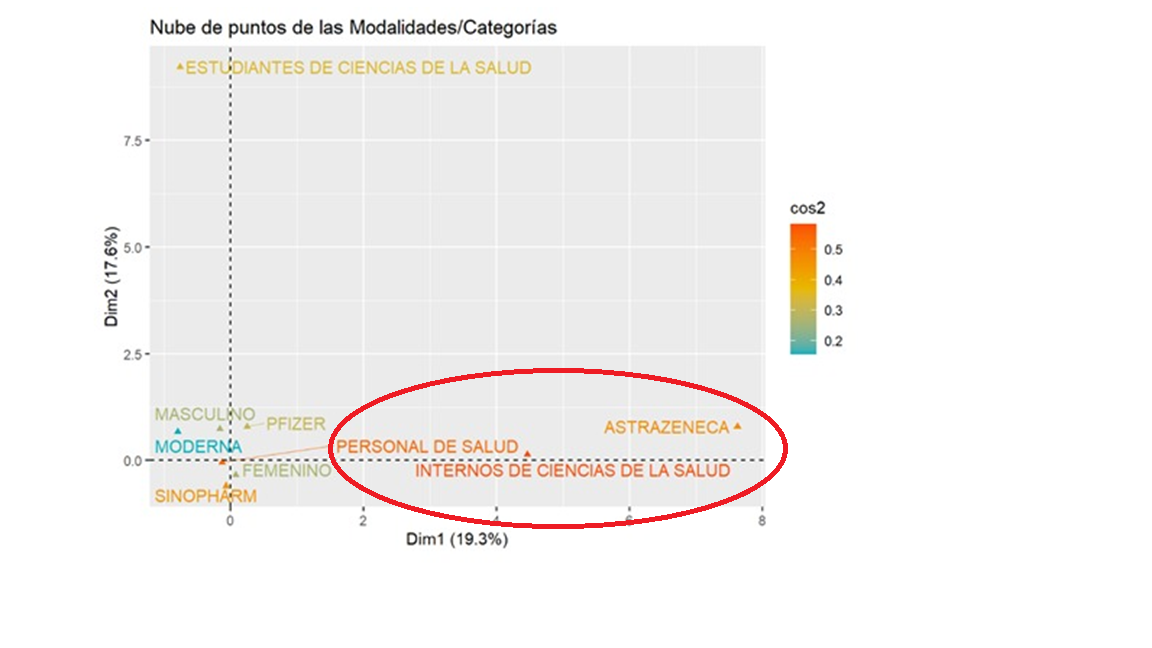}
  \caption{Point cloud of the Modalities/Categories}
  \label{fig:Imagen6}
\end{figure}

The following graph shows that it is easier to observe the categories that are best represented by dimension 1: health sciences interns, health personnel and vaccine As- traZeneca present a better profile. \autoref{fig:Imagen6}\\

\begin{figure}[H]
  \centering
  \includegraphics[width=0.8\textwidth]{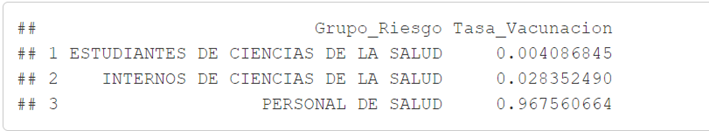}
  \caption{Vaccination Rate}
  \label{fig:Imagen7}
\end{figure}

In the following table, a risk group indicator was identified as Health Sciences Students: The vaccine rate for this group is approximately 0.41\%.This could indicate that a small percentage of health science students have been vaccinated thus far. Health Science Interns: The vaccination rate for this group is approximately  2.83 \%.Compared to health science students, there appears to be a higher proportion of vaccinated interns. Health personnel: The vaccination rate for this group is significantly higher, at approximately 96.76 \%.This suggests that the vast majority of health personnel have been vaccinated. \autoref{fig:Imagen7}\\\\

\section*{Application}

\begin{figure}[H]
  \centering
  \includegraphics[width=0.8\textwidth]{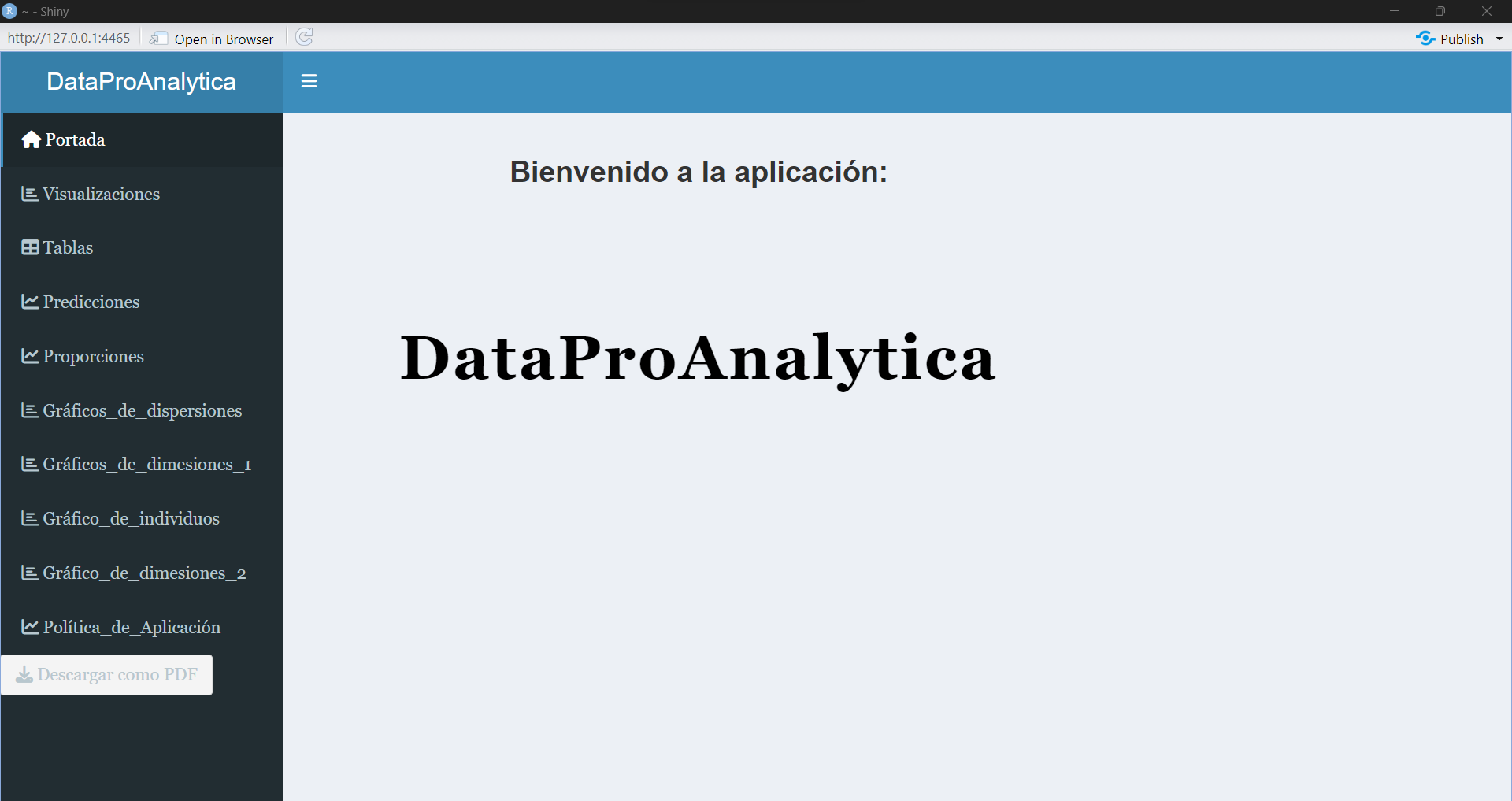}
  \caption{Application}
  \label{fig:apli}
\end{figure}

DataProAnalytica is a data science application that uses multiple correlation analysis, data visualization, and prediction generation. Its versatility promises applications in various databases and research fields. Highlighting its analysis of the vaccination rate in health professionals, this tool will become a fundamental resource to improve strategies and policies in different fields of study \autoref{fig:apli}. The application can be found at the following link:\href{https://tyxfib-luz0bella-valenzuela0narvaez.shinyapps.io/DataProAnalyticaTess/}{click here}

\begin{multicols}{2} % Inicio de las dos columnas
\section{Discussion and Conclusion}
In this study, we present DataProAnalytica, an application designed to analyze vaccination in MINSA health professionals \cite{Los2021}.
Using the method of mcu- piltleorrespondence analysis, we explore the vac- cination rates and the advantages of data visu- alization and analysis offered by the application. DataProAnalytica allows data to be represented in graphs and tables, facilitating the analysis and interpretation of the information, as well as providing tools for prediction and estimation of vaccination- related proportions. The results obtained from the multiple correlation analysis revealed significant relationships between the variables manufacturer and health professionals. Likewise, significant relationships were found between the categories within the aforementioned variables, as well as a significant relationship between the sex variable and health professionals. These relationships were observed mainly in the first dimension of the analysis. In the second dimension, it was found that aztrazeneca category within the manufacturer variable had an estimate of 3.5549 compared to the reference category (base category), with a p value of 0. This indicates that there is a statistically significant difference in the dependent variable between this category and the base category. These findings suggest that the variables and categories analyzed have a significant relationship with the base category significant relationships with each other. However, it was also observed that some categories did not show a significant relationship with individuals, which indicates that not all variables and categories have a relevant influence on the vaccination of health professionals. Regarding the proportion of the vaccination rate, it was found that 97 \%  of the data represent mainly health professionals who have been vaccinated. This is consistent, given that they are the first front in the fight against the highly contagious virus. The remaining 3\% correspond to health science interns, while health students represent a small part of the population. It is important to keep in mind that, due to their student status, they are considered in the same category as the rest of the general population and, therefore, they receive their regular vaccinations regularly, unlike health professionals, who receive more frequent and stronger vaccinations. In summary, DataPro- Analytica is an analysis tool for immunization in MINSA health professionals that uses the method of multiple correspondence analysis. It allows visualization and analysis of data, providing a deeper understanding of the relationships between variables and categories. As future updates are developed, the application will become versatile and can be adapted to different datasets, meeting the needs of users in various fields and industries. This project is only the beginning of a broader development, with the goal of improving the functionality and usefulness of DataProAnalytica in future research. This study provides valuable information on vaccination in health professionals, identifying significant relationships between variables and categories. These results have the potential to improve vaccination strategies and ensure adequate coverage for this crucial group in the fight against infectious diseases \cite{2}.
However, it should be kept in mind that the sample was limited to MINSA health personnel and that the DataProAnalytica application is at an early stage, requiring further improvements and testing to ensure its accuracy and optimal functionalityy \cite{3}.
\section*{Acknowledgments}
We are grateful to the School of Engineering and Computer Science for their valuable support in this study, to the computational statistics course for providing us with fundamental knowledge, and to engineer Fred Torres Cruz for his inspiration and valuable support. His contribution has been fundamental for the success of this project.

\FloatBarrier
\textbf{}

\bibliographystyle{ieeetr}
\bibliography{export.bib}
\end{multicols} % Fin de las dos columnas

\end{document}